\def\mnras{MNRAS}
\def\apj{ApJ}
\def\nature{Nature}

\def\pr{Phys. Rev.}
\def\araa{ARA\&A}

\documentstyle[11pt,aas2pp4]{article}
\begin{document}
\title{Self-Organized Criticality in Compact Plasmas}
\author{Ran Sivron\footnote{Grand Valley State University, Allendale, Michigan, 4940, Email:sivronr@gvsu.edu.}}
\begin{abstract}
Compact plasmas, that exist near black-hole candidates and in gamma ray bursts sources, commonly exhibit self-organized non-linear behavior. A model that simulates the non-linear behavior of a compact radiative plasma is constructed directly from the observed luminosity and variability. The simulation shows that such a plasma self-organizes, and that the degree of non-linearity as well as the slope of the power density spectrum increases with compactness. The simulation is based on a cellular automaton table that includes the properties of the hot (relativistic) plasma, and the magnitude of the energy perturbations. The plasma cools, or heats up, depending on whether it releases more or less energy than that of a single perturbation. The energy released depends on the plasma density and temperature, and the energy of the perturbations. Strong perturbations may cool the previously heated plasma through shocks and/or pair creation. 

New observations of some active galactic nuclei and gamma ray bursts 
are consistent with the simulation. 
\end{abstract}
\section{INTRODUCTION}
Large amplitude X-ray and $\gamma$-ray variability from active galactic nuclei (AGNs), black hole binaries (BHBs), and gamma ray bursts (GRBs) are thought to be the signatures of hot (relativistic) plasma (Blandford 1990, hereafter B90, and references therein, Mushotzky, Done \& Pounds 1993, hereafter MDP93, Fishman \& Meegan 1995, hereafter FM95). Many of these sources exhibit rapid large amplitude variability (MDP93, Green, McHardy \& Lehto 1993, hereafter GML93). The combination of rapid large amplitude variability and high emissivity is thought to be evidence for compact sources (Fabian 1992, B90, MDP93, FM95). 

Roughly half of AGNs and BHBs exhibit $1/f^\alpha$ decline in their power density spectrum (PDS), where $1< \alpha < 2$ (Press 1978, MDP93, McHardy 1988, GML93, Ulrich et al. 1997). Some GRBs also exhibit non-linear oscillations superposed on the general power law decay (Meredith, Ryan \& Young 1995, hereafter MRY95). Recently several time series analysis methods were developed to distinguish between PDS with linear origins and PDS with non-linear origins (e.g., Kaplan \& Glass 1995, hereafter KG95, Vio et al. 1992, hereafter V92, Scargle 1990). By using some of these methods evidence of non-linearity in some AGNs and GRBs was found (V92, Boller et. al. 1997, Leighly \& Obrian 1997, MRY95, Yuan et al. 1996). Some of these light curves were also found to be self-similar
(Scargle, Steiman-Cameron \& Young 1993, hereafter SSY93,  McHardy \& Czerny 1993). 

Several authors have borrowed cellular automaton (CA) models from other scientific fields, to simulate the non-linear behavior of  AGNs, BHBs and GRBs. CAs are tables of rules that describe how to model a non-linear system with identical elements. CAs are commonly used when the partial differential equations for a non-linear system are not easily solvable (Jackson 1989, hereafter J89, KG95). Using a CA Mineshige, Takeuchi \& Nishimori (1994, hereafter MTN94), assumed that the accreting material in BHBs is in the form of an accretion disk, and that the avalanches are analogues to those in the self organized criticality (SOC) CA for sand piles (Bak, Tang \& Wiesenfeld 1988, hereafter BTW88). SSY93 assumed that the material in BHBs is in the form of a ring, as in the dripping hand model of Crutchfield \&  Kaneko (1988). Stern \& Svensson (1997, hereafter SS97) suggested that a pulse-avalanche CA may be appropriate for an expanding "fireball" in GRBs if there is magnetic turbulence due to an unknown instability. These models are motivated by similarity to simulations from other fields, and by the 
{\it{assumed geometry}} of  some astrophysical point sources. The aim of this {\it {Letter}} is to construct an independent CA directly from the {\it{accepted}} physical conditions in the plasma. 

The physical conditions in compact plasma with perturbations and the CA table of rules are described in section \S 2. In section \S 3 it is shown that the plasma indeed evolves to a self-organized critical state for a wide range of emission rates and that the emerging power-law spectrum is of the non-linear $1/f^\alpha$ variety. In section \S 4, the last section, the expected observational ramifications and conclusions are discussed. 
\section{CA MODEL FOR COMPACT PLASMAS WITH PERTURBATIONS}
In many AGNs, BHBs and GRBs there are compact plasmas with nonlinear processes. The energy lost to emission or added to the plasma on a doubling time scale is a significant fraction of the total energy (Sivron 1995, hereafter S95, Piran 1995, hereafter P95). The introduction of a large energy density perturbation into the plasma should result in an increase of emission if the energy is efficiently radiated. For example, emission will rise due to a pair-runaway process in which the increased number of leptons results in efficient Compton cooling.

Post-shock pair cascades are readily produced in compact sources with strong pertrubations (Sivron, Caditz \& Tsuruta, hereafter SCT96). Strong perturbations, in which the perturbation excess density is of the order of the average density in the plasma are known to form shocks (Landau \& Lifshitz 1987). This holds true for relativistic fluids and hot collisional plasma (Taub 1949,  Iwamoto 1989, hereafter I89). In many situations the time scale for perturbation steepening to shocks is small. This is known to be true for compact accreting sources (Papaloizou \& Pringle 1984, Narayan 1991), and is conjectured for GRBs (P95). In such cases one may model all strong super-sonic perturbations as effective radiative shocks that can dominate the light curve (SCT96).
The same strong perturbations may result in very little radiation, when the plasma temperature is too low for pairs to be effectively created. In such a case the perturbation energy would heat up the plasma. All the perturbations are nonlinear in the sense that a strong perturbation significantly changes the plasma temperature and speed of sound, which then determine the output due to the next perturbation. 

In the simulations in this paper it is assumed that non-linear strong perturbations are produced in the source. The input for the simulations is a perturbation moving with input velocity $U_i$, and the output is the luminosity $L_i$, where the number of intervals is $N$, and the running index is $1,2,...,i,...,N$. The simulations follow the CA rules of table 1. The simulation is non-linear in the sense that parameters in cells $i+1$, $i+2$ etc., depend on the output $L_i$. The CA rules and the parameters in it are described in the next three paragraphs.

For demonstrative purposes numbers which are appropriate for a typical source, the active nucleus of a Seyfert I galaxy, are used. The emitting source of size $X= 5 R_{Sch} = 1.5 \times 10^{13}$cm, appropriate for a central black hole of mass $M=10^7 M _{\odot}$,  and the accretion rates relative to the Eddington accretion rate are $\dot{m}=0.1, 0.5, 1.0, 1.5$. Here $R_{Sch}=2GM/c^2$ is the Schwarzchild radius of the putative central black hole with mass $M$, $\dot{m}=\dot{M}/\dot{M_{Edd}}$, $\dot{M}$ is the accretion rate, $\dot{M_{Edd}}=4\pi G M m_H/ (c \sigma_T)$ is the Eddington accretion rate, and $m_H$ and $\sigma_T$ are the hydrogen mass and the Thompson cross section respectively (B90). 

The parameters for the simulation are the following: The time interval is $\Delta$, and $\Delta=1$ second for  the case of Seyfert Is. The total mass of the plasma $m=3\times 10^{24}$ gram, was chosen so that, using the plasma deflection length with electron temperature $\sim 3\times 10^9$K, the plasma deflection length is smaller than $X$, and the plasma is collisional (I89). The total energy of the plasma, not including rest mass, is $E_i$ and the total temperature of the plasma is $T_i=E_i/(1.5 n_i k)$, where $n_i=3m/(m_H 4\pi X^3)$,  $k$ is the Bolzman constant and $m_H$ is the average mass of an atom.  The mass of each perturbation is $m_i={\dot{m} \Delta}$, and the kinetic energy of a perturbation is $(\delta E)_i = [0,(\delta E)_{max}]$ with equal probability\footnote{ The distribution is the result of virial radiative pressure and gravity that produce super and sub-Keplerian speeds for perturbation of  large enough size (S95, SCT96).}. $(\delta E)_{max}$ is the Keplerian energy of perturbation with mass $m_i$ at $5 R_{Sch}$. Using the usual special relativistic expression, the input speed of each perturbation is $U_i=c\sqrt{1-(m_i c^2/(\delta E)_i)^2}$. The output luminosity is $L_i$, and the efficiency of converting gravitational to radiative energy is $\epsilon$\footnote{ In the case of  accretion sources $\epsilon=0.06$ for a Schwarzchild black hole.}. The initial conditions were set at multiples of $E_1=0.05 m c^2$ (see results, section 3). Energy perturbations with Gaussian distributions were also tried. The results were generally the same (see section 3).

The effective shocks are selected in the following way: The parameters  $T_{i}, n_i, U_i$ are sent to a subroutine that uses the Ranking-Hugoniot relations for a hot collisional plasma to find the post-shock conditions (I89, SCT96). If these conditions are sufficient for pair-cascades, and if the perturbation moves faster than $(c_s)_i$, the shock is considered effective (Svensson 1982, Svensson 1984, BS91, SCT96).  The speed of sound in the plasma, $(c_s)_i$ depends on $T_i$, and is typically slightly larger than $c/3$. There are two alternating CAs: If there are no effective shocks the CA is the second row in the table. If there is an effective shock the third row is the CA. The number of columns in that row is $J=(X/U_i)/\Delta+1$, an integer that determines the number of  subsequent cells still affected by effective shocks. In table 1 the width of row B was selected to be $J=3$ for demonstrative purposes.  

The following scenario demonstrates how the non-linearity of the simulation: If initially there is no effective shock the energy $E_{i+1}$, is probably smaller than $E_i$. In such cases $L_i \Delta / \epsilon$ is much smaller than $(\delta E)_{i+1}$,  because the small non-shock emission is diffusion - dominated (see table 1). As a result the energy and temperature of the plasma in the subsequent intervals increase, as does the speed of sound. In the following interval the probability of exceeding the speed of sound is therefore smaller. If, on the other hand,  there is an effective shock, the energy decreases according to columns $2-4$ in the third row, lowering the speed of sound, and increasing the probability of shocking the plasma in the next interval.
\section{RESULTS}
The light curve, PDS, phase, and auto - correlation function (ACF) of the output from the model in Figs 1a, 1b and 1c resemble the results of BTW88, MTN94, SSY93 and SS97. As in BTW88 self organization is achieved independently of initial conditions, because
the system quickly evolves to a state in which perturbations of above average energy shock the plasma. Self organization is not obtained, however, for accretion rates lower than $\dot{m}=0.1$ or larger than $\dot{m}=1.0$. At $\dot{m}< 0.1$ the plasma temperature grows without bound because there are too few large perturbations at supersonic speeds\footnote{In such cases modest magnetic fields are needed for cooling with super Alfvenic perturbations and reconnection events.}.
For $\dot{m}>1.0$ our model was not stable.

In Fig 1a the light curve corresponding with the higher $\dot{m}$ takes longer to reach the organized criticality state because while increasing its temperature the denser plasma loses more energy through small shocks. The minimum doubling time scale is smaller for the higher $\dot{m}$ objects because we increased $\dot{m}$ by lowering the size of the plasma. The shock therefore extracts energy from the plasma over a shorter period, resulting in stronger flares. For accreting sources this corresponds with a smaller compact object.

The PDS exponent in Fig 1b is $\alpha=0.85\pm0.03$ for $\dot{m}=1$ and $\alpha=0.79\pm0.03$ for $\dot{m}=0.5$ (fit not shown). In both cases the PDS includes emission from the time after the critical state has been established, and the subsequent 900 seconds. Energy perturbations with Gaussian distribution of width $0.4 \delta E_{max}$ yielded $\alpha=0.84\pm0.03$ and $0.79\pm0.03$. As expected, for increasing $\dot{m}$  small high frequency shocks are increasingly suppressed as the temperature and associated speed of sound increase just before the more frequent large shocks. With decreasing $\dot{m}$ we get $\alpha\sim 0$, because as the time interval between perturbations grows the system responds linearly to the random perturbations. These results are similar to those in BTW88. In BTW88 a flat  PDS is changed to $1/f^{\alpha}$ due to suppression of short scale avalanches in domains of increasing sizes that were flattened by larger avalanches. In the model presented here the $1/f^{\alpha}$ is due to the suppression of  weak radiative shocks just before stronger shocks. Here the "critical slope" of BTW88, the speed of sound, depends on the previous temperature in the plasma. This model is therefore analogous to a one dimensional BTW88 sand pile with critical angle that depends on the speed and location of previous avalanches.

Correlation on short time scales (in the first few seconds and at around 50-80 seconds) for $\dot{m}=1$ can be seen in the ACF in Fig 1c. 
The enhanced ACF at low time scales makes sense because of the anticipated correlation of perturbations. The peaks are due to the total sum of different average delays due to different conditions in the plasma. The effect of the initial perturbation is lost over time scales $t>X/({\cal{M}}_i (c_s)_i)$, where ${\cal{M}}_i=U_i/(c_s)_i$ is the Mach number. For $\dot{m}=0.5$ there is less overall correlation, but more correlation at times $\sim 80$ seconds (not shown).

Another method by which the non-linear dependence is demonstrated is shown in Fig 2a in which the phase space diagram of the outputs $L_N$ is compared with $L_{N+1}$. With no correlation the path should randomly fill the correlation space, as seen in  Fig 2b for a 100 second delay between $L_N$ with $L_{N+100}$. Despite of the apparent correlation for a 70 second delay the phase space diagram does not show obvious structure (not shown). 
\section{DISCUSSION AND CONCLUDING REMARKS}
The model presented here shows that $1/f^{\alpha}$ non-linear PDS can be created without relying on a specific geometry. One problem with this model is that, in accreting sources, the predicted $\alpha$ is too low. This result can be corrected by adding parametric dimensions. Adding parametric dimensions usually results in larger $\alpha$ (BTW88). We are currently working on a simulation with added parameters, one of which is related to the 2D angular momentum transfer --- an essential component for accreting sources (Sivron \& Leighly 1998, hereafter SL98, Sivron \& Tsuruta 1993, hereafter ST93). For {\it{low $\dot{m}$ }} and flat geometry the added parameter is expected to yield results similar to those of MTN94, because the random input and output for each disk cell are correlated. A parameter related to the profiles of radiation events from each shock is expected to yield a smaller $\alpha$, due to relativistic effects (SL98). This parameter is analogous to the time profile of shots in Takeuchi, Mineshige \& Negoro 1995. 
Another problem with the model is that, contrary to observations, there is no correlation on a time-scale longer than $10^{-3}$ seconds for BHBs (Negoro et al. 1994).  This problem will be corrected with more dimensions because $\dot{m}$ will decrease with the less effective angular momentum transfer associated with small non-shocking perturbations,  making the subsequent shocks less effective (SL98).
 
For accreting sources the simulation yields average output temperatures that are lower for high $\dot{m}$. This is because of an increase in emission efficiency with increasing $\dot{m}$. However, when $\dot{m}$ is high the cooling of the post-shock plasma is effective, and a large portion of the post-shock material is cooled to a "cold phase" of temperature $\sim 10^{6 -7}$K. The effect of such matter on observations is an enhancement of the soft X-ray emission. (Guilbert \& Rees 1988, Celotti Fabian \& Rees 1991, Sivron \& Tsuruta 1993, SCT96, Kuncic, Celotti \& Rees 1997). This result is already consistent with observations of narrow line Seyfert galaxies type I (NLSIs) and "regular" Seyfert Is. NLSIs that have the same emission rate of regular Seyferts  have steeper X-ray spectrum, a slightly larger $\alpha$, and probably smaller black hole and larger $\dot{m}$ (Leighly 1997, SL98). 

The correlation of $\dot{m}$, $\alpha$ and enhanced thermal emission is general. The model therefore predicts that energy density perturbations in the {\it{initial}} fireball of GRBs can produce non-linear temporal variations. These variations can then give rise to winds with varying Lorenz factor $\Gamma$ (S95, SL98). The emission can be the result of low $\Gamma$ winds loaded with baryonic matter overtaking the high $\Gamma$ initial winds that slow down as the fireball sweeps up external matter (Rees \& Meszaros 1997). The non-linear variations in this scenario are {\it{frozen in}}, and reveal themselves when the initial fireball expands and becomes optically thin. This scenario is consistent with GRBs observed light curves. 

We acknowledge helpful remarks made by the anonymous referee, and by 
K. Leighly, S.Tsuruta, B. Hinrichs, J.Scargle,  J. Weber, N. Sivron and R. Svensson.

\begin{figure}
\caption{a. The upper light curve corresponds with the higher $\dot{m}=1$, and takes longer to reach the organized criticality state, but the minimum doubling time scale is smaller. The lower curve corresponds with $\dot{m}=0.5$.
b. In the lower box the solid curve represents the PDS. The dashed curve is the fit to the PDS in the case $\dot{m}=1$, with $\alpha=0.86\pm0.03$. In the upper box the solid curve is the phase of the PDS.
c. The auto-correlation function for $\dot{m}=1$ (bold) is compared with the auto-correlation function for the 
more jagged random input, which has a typical exponential decay.
There is an enhancement in correlation in the first few seconds and at around 50-80 seconds.}
\end{figure}
\begin{figure}
\caption{
a.  Phase space diagram with 1 second delay shows a very clear sign of non-linear dependence. The numbers on all axes are in arbitrary units. The order of points is such that for a point $N$ on the lower accumulation line the point $N+1$ is on the upper accumulation line.
b. The same phase relation with 100 second delay shows no dependence at that time interval, corresponding with loss of information over times larger than $X/U_N$.}
\end{figure}
\clearpage
{\bf{Table 1:The CA Rules}}\vskip 12pt

\def\entry#1:#2:#3:#4:#5 {\strut #1&#2&#3&#4&#5\cr}
{\offinterlineskip \tabskip 0pt \halign{%
\vrule\quad\hfil#\hfil\quad\vrule&
\quad\hfil#\hfil\quad\vrule&
\hfil#\hfil\vrule&
\quad\hfil#\hfil\quad\vrule&
\quad\hfil#\hfil\quad\vrule\cr
\vphantom{\vrule height 2pt}&&&&\cr \noalign{\hrule}
\vphantom{\vrule height 2pt}&&&&\cr 
\entry Conditions for shock:L at cell $i$:Effects of $E_i$:Effects of $E_i$:on
\entry ($T_i=T(E_i),cs_i=cs(T_i)$)::on cell $i+1$:on cell $i+2$:$i+3$
\entry :$t=i\Delta$:$t=(i+1)\Delta$:$t=(i+2)\Delta$:$t=(i+3)\Delta$
\vphantom{\vrule height 2pt}&&&&\cr \noalign{\hrule}
\vphantom{\vrule height 2pt}&&&&\cr 
\entry IF $U_i<cs_i^{**}$:$L_i=dif^{\dag}$:$E_{i+1}=E_i+(\delta E)_i$:no:no
\entry no effective shock::$\quad\;\;-L_i*\Delta/\epsilon$:effect:effect
\vphantom{\vrule height 2pt}&&&&\cr \noalign{\hrule}
\vphantom{\vrule height 2pt}&&&&\cr 
\entry If $U_i>cs_i^{**}$ :$L_i=0.5\epsilon E_i$:$E_i+(\delta E)_i/J-$:$E_{i+1}+(\delta E)_i/J-$:$E_{i+2}+(\delta E)/J-$
\entry shock is effective:$\quad\;/(J+1)^{\dag\dag}$:$\quad\;\;{{L_{i}*\Delta}/\epsilon}$:$\quad\;\;{{L_{i}* \Delta}/ \epsilon}$:$\quad\;\;{L_{i}*\Delta/\epsilon}$ 
\vphantom{\vrule height 2pt}&&&&\cr \noalign{\hrule}}}

\noindent
$^*$ The approximate expressions are: $T_i\sim 2E_i/3k$, $cs_i\sim \sqrt {2kT_i/m}$. (The fully relativistic expressions are from I89.)

\noindent
$^{**}$ More accurately, effective shocks are only those shocks in which $U_i>cs_i$ and copious pair production is achieved (see SCT 96).

\noindent
$^{\dag}$The results of White \& Lightman (1989) for Comptonized bremsstrahlung emission in hot two - temperature plasmas for a given $\dot{m}$ are used for the diffusive emission $dif$. Hot two-temperature plasmas are expected for $\dot{m} \sim 1$ (Rees et. al. 1982, Narayan \& Yi 1994).

\noindent
$^{\dag \dag}$The $0.5$ parameter: due to assumption that half of the energy in the plasma at time $t_{i}$ is radiated away following a shock (see text). 
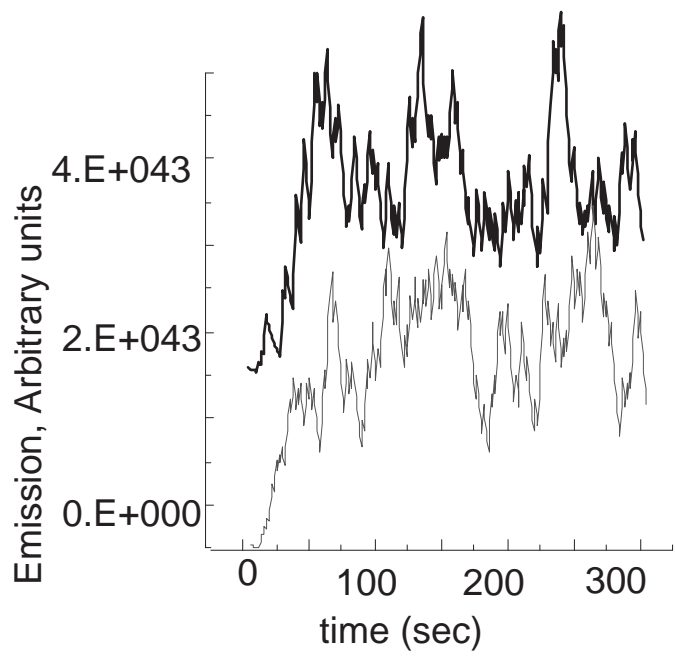
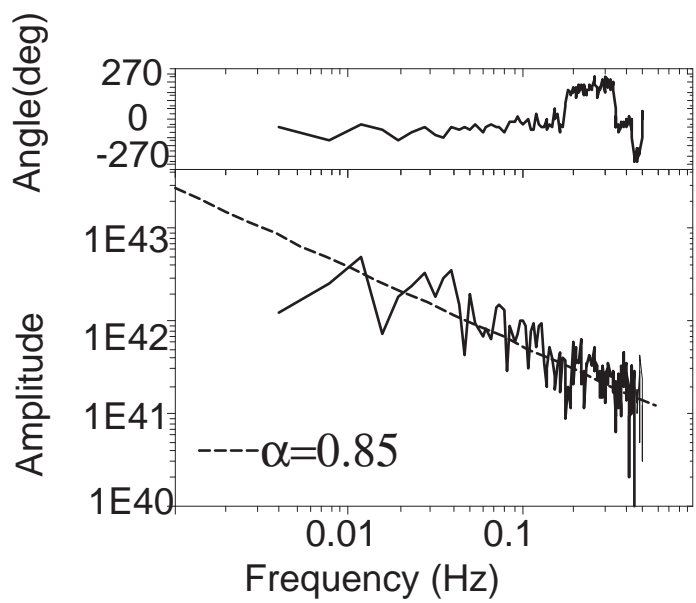
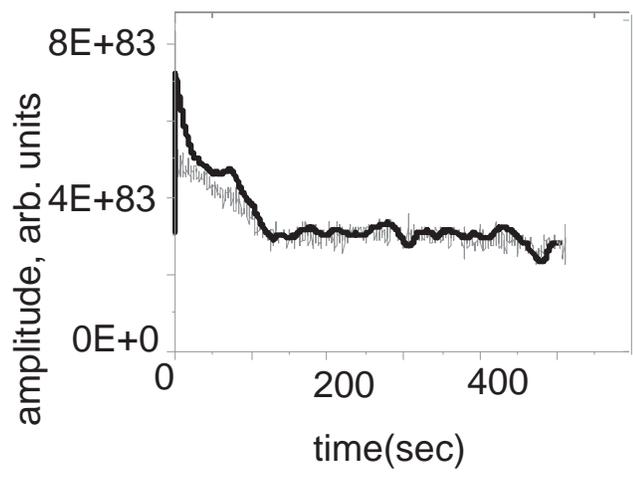
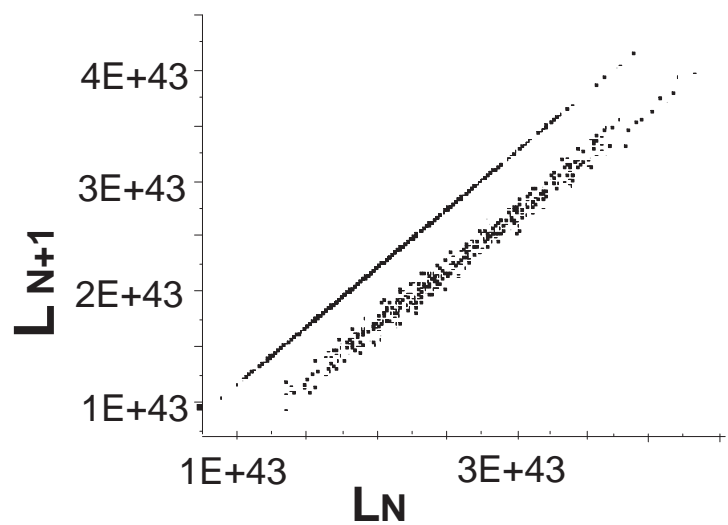
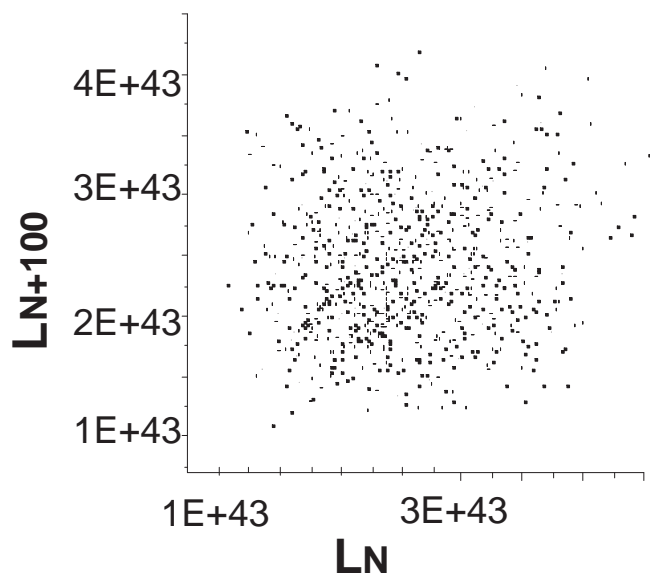
\end{document}